\documentclass[acmsmall,nonacm]{acmart}
\setcopyright{none}
\renewcommand\footnotetextcopyrightpermission[1]{}
\usepackage[utf8]{inputenc}
\usepackage{natbib}
\usepackage{amsmath}
\usepackage{amsfonts}
\usepackage[frozencache=true,cachedir=minted-cache]{minted}
\setminted{fontsize=\small}
\usepackage{newunicodechar}
\usepackage{xspace}
\usepackage{url}
\usepackage{hyperref}

\newcommand{\hs}[1]{\mintinline{haskell}{#1}}

\begin{document}

\title{Toward Hole-Driven Development in Liquid Haskell}

\author{Patrick Redmond}
\affiliation{\institution{University of California, Santa Cruz} \country{USA}}
\author{Gan Shen}
\affiliation{\institution{University of California, Santa Cruz} \country{USA}}
\author{Lindsey Kuper}
\affiliation{\institution{University of California, Santa Cruz} \country{USA}}

\begin{abstract}
  Liquid Haskell is an extension to the Haskell programming language that adds support for \emph{refinement types}: data types augmented with SMT-decidable logical predicates that refine the set of values that can inhabit a type.  Furthermore, Liquid Haskell's support for \emph{refinement reflection} enables the use of Haskell for general-purpose mechanized theorem proving.  A growing list of large-scale mechanized proof developments in Liquid Haskell take advantage of this capability.  Adding theorem-proving capabilities to a ``legacy'' language like Haskell lets programmers directly verify properties of real-world Haskell programs (taking advantage of the existing highly tuned compiler, run-time system, and libraries), just by writing Haskell.  However, more established proof assistants like Agda and Coq offer far better support for interactive proof development and insight into the proof state (for instance, what subgoals still need to be proved to finish a partially-complete proof).  In contrast, Liquid Haskell provides only coarse-grained feedback to the user --- either it reports a type error, or not --- unfortunately hindering its usability as a theorem prover.

In this paper, we propose improving the usability of Liquid Haskell by extending it with support for Agda-style \emph{typed holes} and interactive editing commands that take advantage of them.  In Agda, typed holes allow programmers to indicate unfinished parts of a proof, and incrementally complete the proof in a dialogue with the compiler.  While GHC Haskell already has its own Agda-inspired support for typed holes, we posit that typed holes would be especially powerful and useful if combined with Liquid Haskell's refinement types and SMT automation.  We discuss how typed holes might work in Liquid Haskell, and we consider possible implementation approaches and next steps.
  
\end{abstract}

\maketitle

\section{Introduction}

\emph{Refinement types}~\citep{rushby-predicate-subtyping, xi-array-dependent} are data types augmented with logical predicates, called \emph{refinement predicates}, that restrict, or \emph{refine}, the set of values that can inhabit a type.  Depending on the expressivity of the language used for the refinement predicates, programmers can specify rich program properties using refinement types, sometimes at the expense of the decidability of type checking.  \emph{Liquid types}~\citep{rondon-liquid-types, vazou-lh} are refinement types that avoid such undecidability by restricting refinement predicates to a decidable logic that can be checked by an off-the-shelf SMT solver.
Liquid types are beginning to make their way into general-purpose, industrial-strength programming languages through tools such as \emph{Liquid Haskell}~\citep{vazou-lh}, an extension to the Haskell programming language that adds support for liquid types.

Liquid Haskell's support for \emph{refinement reflection}~\citep{vazou-refinement-reflection, vazou-theorem-proving-for-all} lets programmers use refinement types to specify arbitrary ``extrinsic'' properties that can relate multiple functions, and then prove those properties by writing Haskell programs to inhabit the specified refinement types.
Thanks to refinement reflection and Liquid Haskell's \emph{proof combinators}~\citep{vazou-refinement-reflection, vazou-theorem-proving-for-all}, a Liquid Haskell programmer can leverage the Curry-Howard correspondence and prove arbitrary (undecidable) properties by \emph{inhabiting types with programs}, just as she might do with a dependent-types-based proof assistant like, say, Agda~\citep{norell-agda}, thus ``turning Haskell into a theorem prover''~\citep{vazou-refinement-reflection} and enabling general-purpose mechanized theorem proving in Liquid Haskell.  A growing number of large-scale mechanized proof developments in Liquid Haskell~\citep{vazou-tale-two-provers, parker-lweb, liu-verified-rdts} make use of this capability.

As previous work~\citep{vazou-tale-two-provers, parker-lweb} has discussed, deciding to use Liquid Haskell for mechanized proofs involves a trade-off.  On one hand, with the addition of refinement reflection, Liquid Haskell checks many of the boxes on a modern proof engineer's wish list: Curry-Howard-style proving by programming, SMT automation, and above all, seamless integration with an existing, general-purpose language (with no additional code extraction step required).  Attaching theorem-proving capabilities to a ``legacy'' language like Haskell lets programmers directly verify properties of real-world Haskell programs (taking advantage of the existing highly tuned compiler, run-time system, and libraries), just by writing Haskell~\citep{vazou-dissertation}.
On the other hand, traditional Curry-Howard-based proof assistants like Agda~\citep{norell-agda} or Coq~\citep{bertot-coq} not only rest on a firm theoretical foundation that Liquid Haskell arguably lacks, but they also currently offer far better support for \emph{interactive} proof development and insight into the proof state (for instance, what subgoals still need to be proved to finish a partially-complete proof).  Liquid Haskell provides only coarse-grained feedback to the programmer: either it reports a type error, which means there is still more work to do to complete a proof, or it does not, which means the proof is done.  While such coarse-grained feedback might be fine for intrinsic refinement type specifications or short proofs, the increasing use of Liquid Haskell for large-scale proof development motivates the need for better interactive proof development tools.

In Agda, the interactive proof development experience is enabled in large part by a feature called \emph{typed holes}.  Typed holes allow Agda programmers to indicate parts of a proof that they need help with filling in.  In an interactive environment (such as an IDE), the programmer can use holes to interact with the proof checker by, for example, asking for the type expected by the hole (that is, the proposition that needs to be proven to fill in that hole).  Additional IDE commands can assist the programmer in filling in holes~\citep{agda-docs-typed-holes}.  For example, if the programmer has partially filled a hole with an expression whose \emph{return} type matches the type expected by the hole, then an IDE command can generate new holes indicating the types of the arguments of the expression.  For a hole on the right-hand side of a definition, another IDE command can case-split on a pattern variable and generate new holes for each case.
The result is that, even though Agda favors a style in which programmers explicitly write proof terms (compared to, for instance, proof assistants like Coq or Isabelle~\citep{wenzel-isabelle}, in which programmers more typically use a tactic language or proof automation to indirectly construct proof objects~\citep{ringer-qed}), much of the effort of this explicit proof-writing in Agda is carried out with helpful machine assistance.

In this paper, we propose improving the interactivity of Liquid Haskell by extending it with support for Agda-style typed holes and interactive editing commands that take advantage of them.  In fact, vanilla GHC Haskell already has support for typed holes~\citep{ghc-typed-holes}, a feature that was itself inspired by Agda.  However, we hypothesize that typed holes could be especially powerful in combination with refinement types and SMT automation.  In the rest of this paper, we give an overview of refinement types and Liquid Haskell (\S\ref{sec:background-lh}), and describe how typed holes might work in Liquid Haskell (\S\ref{sec:mockup}).  We conclude with a discussion of possible implementation approaches and next steps (\S\ref{sec:next-steps}).

\section{Background: Refinement Types and Liquid Haskell}
\label{sec:background-lh}

\emph{Refinement types}~\citep{rushby-predicate-subtyping, xi-array-dependent} let programmers specify data types augmented with logical predicates, called \emph{refinement predicates}, that restrict the set of values that can inhabit the type.  Depending on the expressivity of the language of refinement predicates, programmers can specify rich program properties using refinement types, sometimes at the expense of the decidability of type checking.  Liquid Haskell avoids that problem by restricting refinement predicates to an SMT-decidable logic~\citep{rondon-liquid-types, vazou-lh}.  For example, in Liquid Haskell we could define the type of even integers by refining the Haskell type \hs{Int} using the refinement type
\hs{{ v:Int | v mod 2 == 0 }}, where \hs{v mod 2 == 0} is the refinement predicate and \hs{v:Int} binds the name \hs{v} for values of type \hs{Int} that appear in the refinement predicate.
One could define an analogous refinement type for odd integers, and then write a Liquid Haskell function for adding them:

\begin{minted}{haskell}
type EvenInt = { v:Int | v mod 2 == 0 }
type OddInt = { v:Int | v mod 2 == 1 }

oddAdd :: OddInt -> OddInt -> EvenInt
oddAdd x y = x + y
\end{minted}

\noindent The type \hs{OddInt} of the arguments to \hs{oddAdd} expresses the \emph{precondition} that \hs{x} and \hs{y} will be odd, and the return type \hs{EvenInt} expresses the \emph{postcondition} that \hs{x + y} will evaluate to an even number.  Liquid Haskell automatically proves that such postconditions hold by generating verification conditions that are checked at compile time by the underlying SMT solver, Z3~\cite{de-moura-z3}.  If the solver finds a verification condition to be invalid, typechecking fails.  If the return type of \hs{oddAdd} had been \hs{OddInt}, for instance, the above code would fail to typecheck.

Aside from preconditions and postconditions of individual functions, Liquid Haskell makes it possible to verify \emph{extrinsic properties} that are not specific to any particular function's definition.  For example, the type of \hs{sumOdd} below expresses the extrinsic property that the sum of an odd and an even number is an odd number:

\begin{minted}{haskell}
sumOdd :: x : OddInt -> y : EvenInt -> { _:Proof | (x + y) mod 2 == 1 }
sumOdd _ _ = ()
\end{minted}

\noindent Here, \hs{sumOdd} is a Haskell function that returns a \emph{proof} that the sum of \hs{x} and \hs{y} is odd.  (In Liquid Haskell, \hs{Proof} is a type alias for Haskell's \hs{()} (unit) type, but refined with constraints that express the property we wish to prove.)  Because the proof of this particular property is easy for the SMT solver to carry out automatically, the body of the \hs{sumOdd} function need not say anything but \hs{()}, the sole inhabitant of the unit type.  In general, however, programmers can specify arbitrary extrinsic properties in refinement types, including properties that refer to arbitrary Haskell functions via \emph{refinement reflection}~\cite{vazou-refinement-reflection}.
The programmer can then prove those extrinsic properties by writing Haskell programs that inhabit those refinement types, using Liquid Haskell's provided \emph{proof combinators} --- with the help of the underlying SMT solver to simplify the construction of these proofs-as-programs~\cite{vazou-theorem-proving-for-all, vazou-refinement-reflection}.

Liquid Haskell thus occupies a unique position at the intersection of SMT-based program verifiers such as Dafny~\cite{leino-dafny}, and proof assistants that leverage the Curry-Howard correspondence such as Coq and Agda.  A Liquid Haskell program can consist of both application code like \hs{oddAdd} (which runs at execution time, as usual) and verification code like \hs{sumOdd} (which only ``runs'' at compile time), but, pleasantly, both are just Haskell programs, albeit annotated with refinement types.  Being based on Haskell enables programmers to gradually port code from vanilla Haskell to Liquid Haskell, adding richer specifications to code as they go.  Furthermore, verified Liquid Haskell libraries can be used directly in arbitrary Haskell programs, letting programmers take advantage of formally-verified components from unverified code written in an industrial-strength, general-purpose language.
Finally, unlike Coq or Agda, which require an executable implementation to be \emph{extracted} from the code written in the proof assistant's vernacular language, Liquid Haskell enables proving properties both \emph{in} and \emph{about} the code to be executed, resulting in immediately executable verified code with no need for a further extraction step.

\subsection{Why do Liquid Haskell programmers need typed holes?}

Because the proof of \hs{sumOdd} above can be carried out automatically by the SMT solver, there is no need to write anything but \hs{()} as the body of \hs{sumOdd}.  If SMT automation makes proofs so easy to write, why would Liquid Haskell programmers need typed holes, anyway?  Our answer is that, thanks to refinement reflection, large-scale proof development~\citep{vazou-tale-two-provers, parker-lweb, liu-verified-rdts} is now possible in Liquid Haskell, which in turn motivates the need for better machine support in proof construction.

For example, \citet{liu-verified-rdts} used Liquid Haskell to develop a framework for programming distributed applications based on \emph{conflict-free replicated data types} (CRDTs)~\citep{shapiro-crdts}.  CRDTs are data structures whose operations must satisfy
certain mathematical properties that can be leveraged to
ensure \emph{strong convergence}~\citep{shapiro-crdts},
meaning that replicas are guaranteed to have equivalent states
given that they have received and applied the same \emph{unordered} set of update
operations.  For example, consider a data structure representing the contents of a shopping cart, replicated across data centers in Seattle, Frankfurt, Mumbai, and Tokyo for fault tolerance and data locality.  Due to network partitions and message latency, updates to the cart's contents may arrive in an arbitrary order at each data center, but strong convergence ensures that under reliable message delivery assumptions, the replicas will eventually agree.
\citet{liu-verified-rdts}'s framework required proofs that CRDT operations commute.  While these proofs were trivial for some simple CRDTs, more sophisticated ones required on the order of thousands of lines of Liquid Haskell (and hours of solver time)~\citep[Table~3]{liu-verified-rdts}, constituting one of the largest Liquid Haskell proof developments to date.  \citeauthor{liu-verified-rdts} describe the effort as ``strenuous'', highlighting the need for tools that enable visibility into the in-progress proof state.

\section{What could we do with typed holes in Liquid Haskell?}
\label{sec:mockup}

As a simple mock-up of how typed holes might work in Liquid Haskell, consider a \hs{listLength} function that we want to prove has the same behavior as Liquid Haskell's built-in \hs{len}.  One way to state this property is as an intrinsic property on the type of \hs{listLength} itself, as follows:

\begin{minted}{haskell}
listLength :: xs:_ -> { v : Nat | v == len xs }
listLength []     = _0
listLength (y:ys) = 1 + _1
\end{minted}

Here, \hs{_0} and \hs{_1} are holes, indicating that the programmer needs help
filling in these parts of the code.  Given the refinement type annotation on
\hs{listLength}'s return type, together with the context that \hs{xs == []} in
this case, we could expect that the type checker would be able to tell us that
the type of the first hole is \hs{{ v : Nat | v == len [] }}.
The second hole could have type \hs{{ v : Nat | 1 + v == len (y:ys) }}, but the
type checker should be able to take advantage of the underlying SMT solver's
automated reasoning about linear arithmetic to conclude
that the type of the second hole is \hs{{ v : Nat | v == len (y:ys) - 1 }}.
So we have:

\begin{minted}{haskell}
_0 : { v : Nat | v == len [] }
_1 : { v : Nat | v == len (y:ys) - 1 }
\end{minted}

\noindent which can be further simplified to

\begin{minted}{haskell}
_0 : { v : Nat | v == 0 }
_1 : { v : Nat | v == len ys }
\end{minted}

\noindent at which point we know we need \hs{0} and \hs{listLength ys} for \hs{_0} and \hs{_1} respectively.

The preceding example shows how a programmer might use typed holes for writing code with intrinsic refinement type specifications, but how about for writing extrinsic proofs?  Instead of giving a refinement type to \hs{listLength}, we can express the property that relates \hs{listLength} and \hs{len} extrinsically, as follows:

\begin{minted}{haskell}
listLength :: [a] -> Nat
listLength [] = 0
listLength (y:ys) = 1 + listLength ys

listLengthProof :: xs:_ -> { _:Proof | listLength xs == len xs }
listLengthProof = _0
\end{minted}

\noindent The body of \hs{listLengthProof} is a hole, written as \hs{_0}.  When we attempt to compile this code, since the property we are trying to show relates the return values of \hs{listLength} and \hs{len}, Liquid Haskell might start by suggesting a case split on one of the functions:

\begin{verbatim}
Found hole `_0’ of type `xs:[a] -> { _:Proof | listLength xs == len xs }’.
       Consider a case split as in the body of `listLength’.
\end{verbatim}

The case split could be automated with a keystroke, or manually completed.  In either case, the result is a partially completed proof with two holes:

\begin{minted}{haskell}
listLengthProof :: xs:_ -> { _:Proof | listLength xs == len xs }
listLengthProof [] = _0
listLengthProof (y:ys) = _1
\end{minted}

\noindent The first hole, \hs{_0}, has expected type \hs{{ _:Proof | listLength xs == len xs }}.  In \hs{_0}'s context, Liquid Haskell could substitute \hs{[]} in place of \hs{xs} in \hs{listLength xs} and then evaluate the call to \hs{listLength} to arrive at \hs{0}, with similar steps to evaluate \hs{len xs} to \hs{0}, to complete the proof.  This process could be automated after Liquid Haskell encounters the hole by checking whether \hs{()} completes the branch's proof in the background and then issuing a message:

\begin{verbatim}
Found hole `_0' of type `{ _:Proof | xs == [] && listLength xs == len xs }'.
       This can be completed with `()'.
\end{verbatim}

\noindent The replacement of \hs{_0} with \hs{()} could again be automated with a keystroke, or manually completed.  Even better, completing branches with \hs{()} could be folded into automated case-splitting such that each branch is generated with either \hs{()} or a new hole, according to whether more work is needed.

The second hole, \hs{_1}, has expected type \hs{{ _:Proof | listLength xs == len xs }}.  In \hs{_1}'s context, Liquid Haskell could substitute \hs{(y:ys)} in place of \hs{xs} in \hs{listLength xs} and then evaluate the call to \hs{listLength xs} to arrive at \hs{1 + listLength ys}, with similar steps for \hs{len}, to generate a new goal \hs{{ _:Proof | 1 + listLength ys == 1 + len ys}}. This could be automated after Liquid Haskell encounters the hole and finds that replacing it with \hs{()} does not complete the proof. Liquid Haskell could then issue a message:

\begin{verbatim}
Found hole `_1' of type `{ _:Proof | xs == (y:ys) && listLength xs == len xs }'.
       Conclusion expands to `1 + listLength ys == 1 + len ys',
       which is simplified to `listLength ys == len ys`.
\end{verbatim}

\noindent This message does not suggest any action directly, but it does show us an intermediate goal which if proved would be sufficient to show the overall conclusion \hs{listLength xs == len xs}. Users may recognize this goal as the \emph{inductive assumption}. Replacing \hs{_1} with \hs{listLengthProof ys} could potentially be automated with a keystroke in this simple example, although it is unlikely to be quite this obvious in nontrivial proofs.

\section{Next Steps}
\label{sec:next-steps}

In this section, we consider ways to implement the behavior sketched in \S\ref{sec:mockup} by extending the existing Liquid Haskell implementation, and discuss how the proposed behavior differs from what GHC Haskell's typed holes already offer.

\paragraph{Existing Liquid Haskell implementation.} In 2020, Liquid Haskell transitioned from an external tool called
\texttt{liquid} to one implemented as a GHC plugin~\citep{jhala-lh-plugin}.
Following the GHC build sequence, the \texttt{LiquidHaskell} plugin checks
refinement types after GHC completes its typechecking.
The plugin obtains a refinement type for every top-level binding and component
expression through a combination of annotations and inference.

Existing (unpublished) work on
synthesizing Haskell expressions using Liquid Haskell has already seen the
plugin internals set up to capture holes of the form \hs{_goal}.
When a hole is captured, Liquid Haskell tracks the logical environment at the
location of the hole.
This environment has the potential to be used as a Coq-style proof-writing environment so that
the user can learn what information is missing.
Leveraging this existing work, a path forward might be:
\begin{itemize}
    \item Collect examples of hole placements in nontrivial proofs.
    \item Observe the information tracked in the logical environment at those holes.
    \item Experiment with ways to convey that information concisely to the user in compiler or IDE output.
\end{itemize}
With this foundation, we'll be able to decide how to proceed.
Options include:
\begin{itemize}
    \item Transforming the logical environment with SMT-aided simplification, as
        in the \hs{listLength} and \hs{listLengthProof} examples above.
    \item Completing the interactivity loop from compiler output (including
        on-disk suggestion metadata) to editor commands which trigger automated
        changes to the user's code.
\end{itemize}

\paragraph{Comparison with GHC's typed holes.}

Typed holes in GHC currently communicate the inferred principal type of
the expression in an error message to the user.
This is useful, and has inspired some Haskell programmers to 
advocate for a ``hole-driven development'' style of programming.
However, in the case of the incomplete \hs{listLength} function and the
incomplete \hs{listLengthProof} from \S\ref{sec:mockup}, the principal types are
obvious, and the real utility would come from analysis of the refinement type
context to generate the constrained refinement type and synthesize expressions
that fit.

Recent work~\citep{gissurarson-hole-fits} has extended GHC's typed hole support to combine
the principal type information available from GHC's inference with the types of
in-scope binders, such as those imported from library modules.
This work suggests both binders with exact type matches and also ones with more
general types that fit.
However, for a program like \hs{listLength} above, without taking refinement type
information into account, there may be many completions that are well-typed but behaviorally incorrect. In the case of \hs{listLengthProof}, there is simply not sufficient
information at the level of Haskell types to guide suggestions or synthesis
meaningfully.
In fact, \citet{gissurarson-hole-fits} suggests integration with Liquid
Haskell as a possible solution to the problem of principal types in GHC being too vague to
suggest useful completions.
Our hypothesis is that integrating typed holes into Liquid Haskell will
allow more knowledge of the programmer's intent, expressed in refinement types,
to drive suggestions and code synthesis.

\bibliography{references}

\end{document}